\documentclass[preprint,showpacs,preprintnumbers,amsmath,amssymb]{revtex4}

\usepackage{graphicx}% Include figure files
\usepackage{dcolumn}% Align table columns on decimal point
\usepackage{bm}% bold math
\usepackage[]{ams,amsmath,amssymb,epsfig,color}
\newcommand{\be}{\begin{equation}}
\newcommand{\ee}{\end{equation}}
\begin{document}

%\preprint{JPhysB}

\title{Paul Trap and the Problem of Quantum Stability}
%Manuscript Title:\\with Forced Linebreak}% Force line breaks with \\
%\title{Dynamics of dipolar molecular chain in crossed static: Soliton formation}
%Manuscript Title:\\with Forced Linebreak}% Force line breaks with \\

%
\author{L. Chotorlishvili , K. Nickoladze}
\author{G. Mchedlishvili}
%

 %\email{}
\affiliation{Physics Department of the Tbilisi State
University,Chavchavadze av.3,0128, Tbilisi,Georgia}
\affiliation{Email: lchotor33@yahoo.com}
\date{\today}% It is always \today, today,
             %  but any date may be explicitly specified

\begin{abstract}
This work is devoted to the investigation of possibility of
controlling of ions motion inside Paul trap. It has been shown
that by proper selection of the parameters of controlling electric
fields, stable localization of ions inside Paul trap is possible.
Quantum consideration of this problem is reduced to the
investigation of the Mathieu-Schrodinger equation. It has been
shown that quantum consideration is appreciably different from
classical one that leads to stronger limitations of the values of
the parameters of stable motion. Connection between the problem
under study and the possibility of experimental observation of
quantum chaos has been shown.
\end{abstract}

%Valid PACS numbers may be entered using the \verb+\pacs{#1}+
%command.
\pacs{73.23.--b,78.67.--n,72.15.Lh,42.65.Re}% PACS, the Physics and Astronomy
                             % Classification Scheme.
%\keywords{Suggested keywords}%Use showkeys class option if keyword
                              %display desired
\maketitle

%\section{\label{Sec:Introduction} Introduction}

\section{Introduction}
It is well known that formation of Bose-Einstein condensate is
based on application of miniature trap technology
\cite{Anderson,Davis,Bradley,Jin,Mewes,Anrews}.

Formation of condensates of alkaline atoms turned out to be
possible thanks to unification of new ideas and technologies
developed in other areas of physics. First of all magnetic trap
must be formed in order to confine the condensate.

In spite of this, the traps formed by application of variable
electric fields are the subjects of special interests. The point
is that, the confinement of separate ions in trap for a long time
gives new possibilities in laser spectroscopy.  In addition,
single entrapped ion represents itself a unique object to prove
the fundamental laws of quantum mechanics
\cite{Aoki,Mabuchi,Hood,Raimond,Wineland}. For example, single ion
placed inside Paul trap is applied to realize a quantum gate, and
a chain consisting of many ions placed in linear trap may be
regarded as promising component of future quantum computer
\cite{Turchette}.

The goal of this work is the study of quantum dynamics based on
the application of Paul trap \cite{Paul,Blatt,Neuhauser,Schleich}.

Not going into details that can be found in \cite{Schleich}, we
shall describe shortly necessary information about Paul trap
needed further.

Paul trap is based exceptionally on the application of
sign-variable electric field. Besides, potential formed by
electrodes inside trap has the following form \cite{Schleich}: \be
\phi(r,t)=f(t)(x^2+y^2z^2 -2z^2), \ee where
$$f(t)=\frac{U+V\cos(\omega_{rf}t)}{r_0^2+2z_0^2}.$$ Here, $U$ and $V$ are the amplitudes of
constant and variable fields, $r_0$-radius of circular electrode,
$2z_0$-distance between covers, and $\omega_{rf}=2\pi/T$ frequency
of variable field.

The ion's classical motion in potential (1) is described by: \be
M\ddot{\vec{r}}=-e\nabla\phi(\vec{r},t), \ee where $M$ and $e$ are
mass and charge of the ion. Putting potential (1) into equation of
motion (2) yields:\be
M\ddot{\vec{r}}+2ef(t)(\vec{r}-3\vec{z})=0,\ee where
$\vec{z}=(0,0,z)$.

If we label:\be
\psi=x,~~\varphi=\frac{1}{2}\omega_{rf}t,~~a=\frac{8eU}{M\omega_{rf}^{2}(r_0^2+2z_0^2)},~~
q=\frac{4eV}{M\omega_{rf}^{2}(r_0^2+2z_0^2)}, \ee Eq.(3) may be
rewritten in the form of the Mathieu equation \cite{Schleich}: \be
\frac{\partial^{2}\psi(\varphi)}{d\varphi}+(a+2q\cos(2\varphi))\psi(\varphi)=0.\ee

The Mathieu equation is a well known equation of mathematical
physics \cite{Abramowitz}], and it is usually encountered when
studding the problems of parametric resonance
\cite{Nayfeh,Sagdeev}.

The presence of the areas of stable and unstable motion is a
peculiarity of parametric resonance. The solutions of Eq. (5) may
be as stable as unstable depending on the values of $a$ and $q$.
In case of Paul trap, ion confinement inside a trap corresponds to
stable solution. Because of this, we are interested in the stable
solutions (2) of Eq. (5). Depending on physical situation for
studying motion of ions inside Paul trap classical and quantum
considerations are used. The modern state of affairs in studying
this problem is given in \cite{Schleich}. In spite of the fact
that, classical consideration of this problem is more or less
adequate, quantum-mechanical consideration is far from perfection
\cite{Schleich}.

The goal of this work is to fill up this gap.

In this work we shall try to give more adequate (in our opinion)
quantum mechanical description of motion of ions inside Paul trap.
At the same time we shall use the results obtained by us from the
study of the problem of quantum chaos
\cite{Ugulava,Chotorlishvili,Nickoladze,Chkhaidze}.

\section{Stable Quantum-Mechanical Solutions. The Mathieu-Schrodinger Equation Symmetries}

The Mathieu Eq.(5) in classical consideration , depending on the
values of the parameters $a$ and $q$, has as stable as unstable
solutions. On the plane $(a,q)$, the boundaries between stable and
unstable solutions pass through special curve called the Mathieu
characteristics [17]. Along these lines, to which certain set of
the values of the parameters $(a(q),q)$ correspond, the well known
periodic Mathieu functions $ce_{n}(\varphi,q),~se_{n}(\varphi,q)$
\cite{Abramowitz} are solutions of classical Mathieu Eq.(5).
Periodic solutions are the subject of our special interest, since
periodic solutions corresponds to finite motion, which means
confinement of ion inside Paul trap. It is worth noting that, this
kind of selection of the values of the parameters of $a(q)$ and
$q$ impose limitations on the amplitudes of constant and variable
fields: $U$ and $V$ (see Eq.(4)).

In quantum consideration situation gets considerably complicated.
The point is that, in quantum case Eq. (5) must be considered as
the Mathieu-Schrodinger equation. In this case solution $
\psi(\varphi,q)$ and the value of  parameter $a(q)$ must be
considered as eigenfunction and eigenvalue of quantum-mechanical
equation. Characteristic property of the Mathieu-Schrodinger
equation is the following: eigenfunctions $\psi_{n}(\varphi,q)$
can differ from the solutions of classical Mathieu equation
$ce_{n}(\varphi,q),~se_{n}(\varphi,q)$. Finding of eigenfunctions
$\psi_{n}(\varphi,q)$ of the Mathieu-Schrodinger equation is
difficult problem and the application of group theory is needed.
Interested reader can find details in work \cite{Chotorlishvili}.
In the present work we shall only use the results needed further.

The eigenfunctions of the Mathieu-Schrodinger equation form the
basis of irreducible  representation of  the Mathieu-Schrodinger
equation symmetry group.

As is known [17], periodic solutions of Eq. (5) are given by the
Mathieu functions: \be ce_{2n}(\varphi,q),~ ce_{2n+1}(\varphi,q),
~se_{2n+1}(\varphi,q),~se_{2n+2}(\varphi,q)\ee which setisfy the
normalization condition:
$$\frac{1}{\pi}\int_{0}^{2\pi}\psi_{n}^{2}(\varphi,q)d\varphi=1$$ where $\psi_{n}(\varphi,q)$ means Mathieu
functions (6). To eigenfunctions (6) there corresponds the
eigenvalues (Mathieu characteristics) \be
{a_{2m}(q),~~a_{2m+1}(q),~~b_{2m+1}(q),~~b_{2m+2}(q)}, \ee which
also depend on the parameters $q$.

The properties of symmetry of the Mathieu functions can be
presented in the form of Table (see Table 1 in Ref.
\cite{Chotorlishvili}).

By immediate check it is easy to be convinced that symmetry
properties of the Mathieu functions with respect to four elements
of transformation:
$$ G(\varphi\rightarrow -\varphi)=a, ~~G(\varphi\rightarrow
\pi-\varphi)=b,$$ $$ G(\varphi\rightarrow
\pi+\varphi)=c,~~G(\varphi\rightarrow \varphi)=e $$ form a group.
For this purpose it is enough to test the realization of the
following relations:$$a^{2}=b^{2}=c^{2}=e,$$
$$ab=c,~~ac=b,~~bc=a.$$

Group $G$ contains three elements $a,b,c$ of the second order and
unity element $e$. The group $G$ is isomorphic to the well-known
group of Klein \cite{Hamermesh}.

The group of transformations $G$ is not a simple group since it
contains subgroups. When combined with the unit element, each of
three elements $a,b,c$ forms a subgroup of second
order:$$G_{+}\rightarrow e,b;$$ $$G_{-}\rightarrow e,c;$$ \be
G_{0}\rightarrow e,a. \ee

Thus the Mathieu-Schrodinger equation is characterized by a
presence of three symmetries determined by subgroups of symmetry
group. Depending on the values of parameter $q$, the system can be
in the states characterized by any of the symmetries (8). It is
worth mention that, to the symmetries  $G_{+}$ and $G_{-}$
correspond degenerate states with wave functions [21]:
$$G_{-}\rightarrow
\psi_{2n+1}^{\pm}(\varphi)=\frac{\sqrt{2}}{2}(ce_{2n+1}(\varphi)\pm
ise_{2n+1}(\varphi)),$$ \be
\psi_{2n}^{\pm}(\varphi)=\frac{\sqrt{2}}{2}(ce_{2n}(\varphi)\pm
ise_{2n}(\varphi))\ee and $$G_{+}\rightarrow
\eta_{2n}^{\pm}(\varphi)=\frac{\sqrt{2}}{2}(ce_{2n}(\varphi)\pm
ise_{2n+1}(\varphi)),$$ \be
\eta_{2n+1}^{\pm}(\varphi)=\frac{\sqrt{2}}{2}(ce_{2n+1}(\varphi)\pm
ise_{2n+2}(\varphi)).\ee

To the symmetry group $G_{0}$ correspond nondegenerate states with
wave functions \cite{Chotorlishvili}: \be G_{0}\rightarrow
ce_{2n}(\varphi);~ce_{2n+1}(\varphi);~se_{2n}(\varphi);~se_{2n+1}(\varphi).\ee

In the end of this section, let us sum up some results.

The Mathieu-Schrodinger equation is characterized by a specific
dependence of the spectrum of eigenvalues $a_{n}(q)$  and
eigenfunctions $\psi_{n}(\varphi,q)$ on the parameter $q$ (see
Fig.1). On the plane with the spectral characteristics (so-called
Mathieu characteristics [17]) of the problem, this specific
feature manifests itself in the alternation of areas of degenerate
$(G_{\pm})$ and no degenerate $(G_{0})$ states. The boundaries
between these areas pass through the branch points of energy terms
$a_{n}(q)$.

In case of quantum consideration, the problem of confinement of
ion inside Paul trap, the aforesaid means the following:

In order ion's position to be localized inside Paul trap, it is
necessary that the solutions of the Mathieu-Schrodinger
quantum-mechanical equation to be periodic and as a result finite.
This can be achieved only by proper selection of the values of the
parameters $a(q)$, $q$ corresponding to the Mathieu
characteristics, since along these lines motion is periodic.
According to Eq. (4), this can be done by proper selection of the
parameters of electric field $U$ and $V$. The system with these
kind of selected values of the parameters can be found in the
state characterized by one of the symmetry and it shall be
described by corresponding wave functions (9)-(11) (see Fig.1).

In case of classical consideration the situation is different. Not
only the Mathieu characteristics belong to the areas of stable
motion but also the areas where the Floquet solutions are damped
functions of time \cite{Bateman}.

In case of quantum consideration the areas of damped solutions
belong to forbidden energy zones in semiconductors \cite{Pierls}.
This is characteristic property of quantum case.

\section{Nonstationary case. Stochastic Heating of the System}

Let us assume, that the amplitude of the electromagnetic field $V$
is modulated by a slowly changing field. According to (4), the
influence of modulation can be taken into account by making a
replacement in the Mathieu-Schrodinger equation (5) \be
q(t)=q_{0}+\Delta q\cos \nu t,\ee where $\Delta q$ is the
modulation amplitude expressed in dimensionless units and $\nu$ is
the modulation frequency. In this case, the problem of ion's
motion inside Paul trap is reduced to the problem studied in work
\cite{Nickoladze}. In the mentioned work stochastic absorption of
energy by non stationary chaotic quantum system was studied. Not
going into details that can be found by interested reader in
\cite{Nickoladze}, here we only present the main results related
to given problem.

After making replacement (12), Hamiltonian for the
Mathieu-Schrodinger equation takes the
form:$$\hat{H}=\hat{H_{0}}+\hat{H^{\prime}}(t),$$
$$\hat{H_{0}}=-\frac{\partial^{2}}{\partial \varphi^{2}}+q_{0}\cos
2\varphi,$$\be \hat{H^{\prime}}(t)=\Delta q\cos 2\varphi\cos \nu
t. \ee

We assume that gradual change of $q(t)$ may involve some N branch
points on the left and right side of the separatrix $a(q)=q$, (see
Fig.1).

Simple calculations show that the matrix elements of perturbation
$\hat{H^{\prime}}(t)$ with respect to the wave functions of the
nondegenerate area $G_{0}$ (11) are equal to zero \be <ce_{n}
|\hat{H^{\prime}}(t)| se_{n}>=0.\ee

But in the degenerate areas $G_{+},G_{-}$  \be
H_{+-}^{\prime}=<\psi_{2n+1}|\hat{H^{\prime}}(t)|\psi_{2n+1}^{-}>\neq
0, \ee where $\psi_{2n+1}^{\pm}$are the functions (9),(10)
corresponding to the areas of symmetry $G_{-}$ and $G_{+}$ (see
Fig.1).

Thus the modulation (12) may cause transition between quantum
states of the degenerate areas $G_{+},G_{-}$.

When there is a small dispersion in the values of parameter
$\delta q_{0}$ , as was shown in work \cite{Nickoladze}, mixed
states are formed in the degenerate areas. As a result of this,
the system may be with equal probability of 1/2 in two states
$\psi_{2n}^{+}$ and $\psi_{2n}^{-}$. Consequently, because of
specific dependence of energy terms $a_{n}(q)$  on the parameters
$q$ (see Fig.1), and because of time modulation $q(t)$, the
process is extended to other levels as well. At the same time
absorption of energy by the system and stochastic heating takes
place (see Fig. 8 of work \cite{Nickoladze}).

Something like that must take place in case of ion placed in Paul
trap and kept there by amplitude modulated alternate electric
field.

Thus Paul trap may be used for experimental observation of quantum
chaos. The present theoretical analysis gives some recommendations
concerning the values of dimensionless parameters of the Paul
trap(4): $a\approx6.7\div17.2;~q_{0}\approx12.$

The point is that, the problem of internal rotational motion in
polyatomic molecules is reduced to the Mathieu-Schrodinger
equation model \cite{Flyger}. Therefore as is shown in work
\cite{Chkhaidze}, the present experimental data of infrared
absorption in molecules $C_{2}H_{6}$ \cite{Shimanouchi} are direct
evidence of proposed model to be correct.

\begin{figure}[t]
  \centering
  \includegraphics[width=8cm]{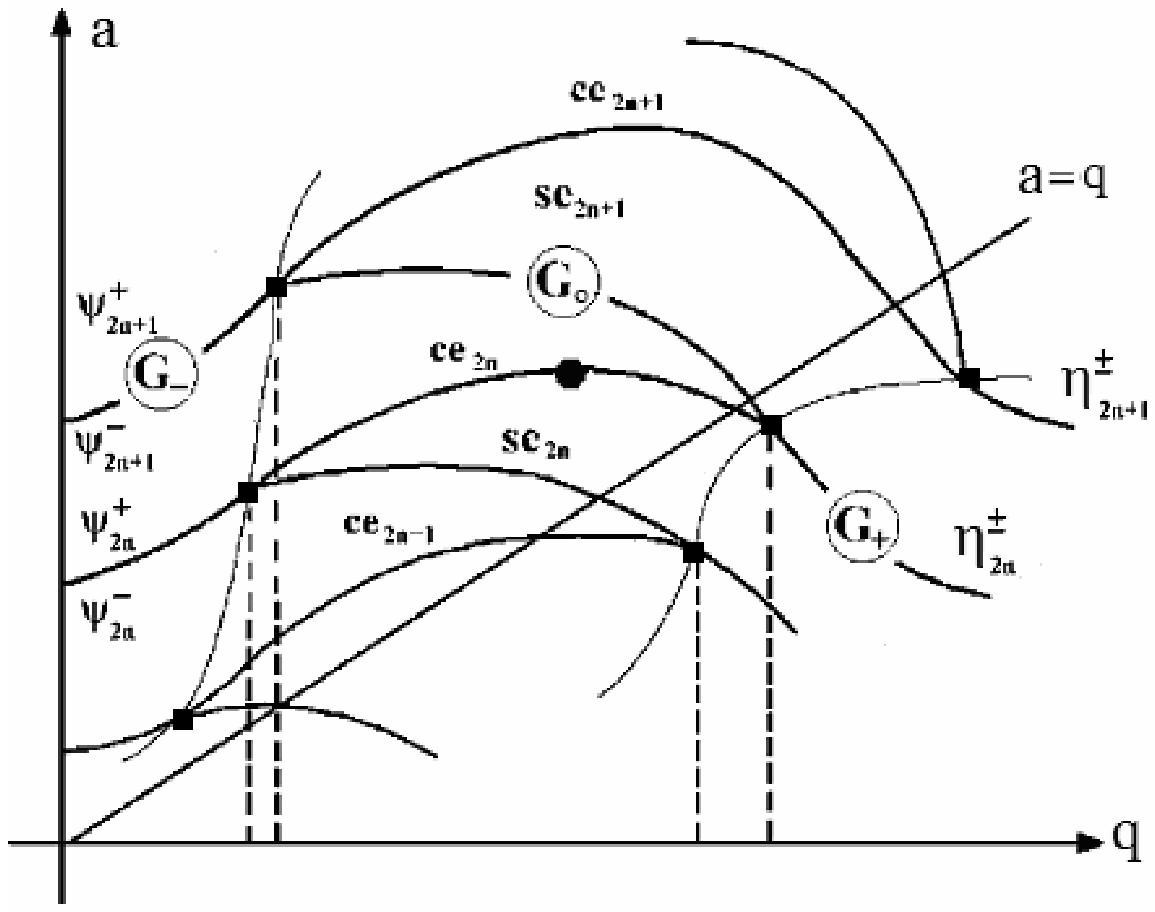}
  \caption{A fragment of the parameter-dependent energy spectrum. Squares denote branch points.}\label{fig:1}
\end{figure}

\end{document}